\documentclass[conference]{IEEEtran}
\IEEEoverridecommandlockouts
\usepackage[figuresright]{rotating}
\usepackage{cite}
\usepackage{amsmath,amssymb,amsfonts}
\usepackage{amsmath}
\usepackage{amsfonts}
\usepackage{stfloats}
\usepackage[ruled,linesnumbered]{algorithm2e}
\usepackage{graphicx}
\usepackage{float}
\usepackage{textcomp}
\usepackage{tablefootnote}
\usepackage{hyperref}
\usepackage{xcolor}
\usepackage{makecell}
\usepackage{adjustbox}
\usepackage{lscape}
\usepackage[ruled,linesnumbered]{algorithm2e}
\usepackage{amsmath}
\usepackage{amsfonts}
\usepackage{bbm}
\usepackage{comment}
\usepackage{enumitem}
\usepackage{graphicx}
\usepackage{makecell}
\usepackage{colortbl}
\usepackage{tikz}
\def\BibTeX{{\rm B\kern-.05em{\sc i\kern-.025em b}\kern-.08em
    T\kern-.1667em\lower.7ex\hbox{E}\kern-.125emX}}
\DeclareMathOperator*{\argmax}{arg\,max}

\usepackage[graphicx]{realboxes}
\definecolor{beaublue}{HTML}{D1E3F1}
\definecolor{beauyellow}{HTML}{F1EFD1}
\usepackage{rotating}
\usepackage{atbegshi}
\newcommand{\method}{\textsc{SynC}\xspace}
\begin{document}

\newcommand{\Yue}[1]{{\color{blue} [Yue comments ``#1'']}}

\title{\method: A Copula based Framework for Generating Synthetic Data from Aggregated Sources}

\author{

\IEEEauthorblockN{Zheng Li}
\IEEEauthorblockA{
\textit{Northeastern University Toronto.}\\
\textit{\&  Arima Inc.}\\
Toronto, Canada \\
winston@arimadata.com}
\and
\IEEEauthorblockN{Yue Zhao}
\IEEEauthorblockA{
\textit{H. John Heinz III College} \\
\textit{Carnegie Mellon University}\\
Pittsburgh, USA \\
zhaoy@cmu.edu}
\and
\IEEEauthorblockN{Jialin Fu}
\IEEEauthorblockA{
\textit{University of Toronto}\\
Toronto, Canada \\
jialin.fu@mail.utoronto.ca}
}

\maketitle

\begin{abstract}
A synthetic dataset is a data object that is generated programmatically, and it may be valuable to creating a single dataset from multiple sources when direct collection is difficult or costly. Although it is a fundamental step for many data science tasks, an efficient and standard framework is absent. In this paper, we study a specific synthetic data generation task called \textbf{downscaling}, a procedure to infer high-resolution, harder-to-collect information (e.g., individual level records) from many low-resolution, easy-to-collect sources, and propose a multi-stage framework called \method (\textbf{Syn}thetic Data Generation via Gaussian \textbf{C}opula). For given low-resolution datasets, the central idea of \method is to fit Gaussian copula models to each of the low-resolution datasets in order to correctly capture dependencies and marginal distributions, and then sample from the fitted models to obtain the desired high-resolution subsets. Predictive models are then used to merge sampled subsets into one, and finally, sampled datasets are scaled according to low-resolution marginal constraints. We make four key contributions in this work: 1) propose a novel framework for generating individual level data from aggregated data sources by combining state-of-the-art machine learning and statistical techniques, 2) perform simulation studies to validate \method's performance as a synthetic data generation algorithm, 3) demonstrate its value as a feature engineering tool, as well as an alternative to data collection in situations where gathering is difficult through two real-world datasets, 4) release an easy-to-use framework implementation for reproducibility and scalability at the production level that easily incorporates new data.
\end{abstract}


\section{Introduction}
Synthetic data is a data object that is artificially created rather than collected from actual events. It is widely used in applications like harmonizing multiple data sources or augmenting existing data. In many practical settings, sensitive information such as names, email addresses, phone numbers are considered personally-identifiable, and hence are not releasable. However, these fields are natural keys to combine multiple data sources collect by different organizations at different times. To overcome this, synthetic data generation becomes a very attractive alternative to obtaining data for practitioners. To efficiently produce high quality data, we study a procedure called downscaling, which attempts to generate high-resolution data (e.g., individual level records) from multiple low-resolution sources (e.g., averages of many individual records). Because low-resolution data is no longer personally-identifiable, it can be published without concerns of releasing personal information. However, practitioners often find individual level data far more appealing, as aggregated data lack information such as variances and distributions of variables. For the downscaled synthetic data to be useful, it needs to be \textit{fair} and \textit{consistent}. The first condition means that simulated data should mimic realistic distributions and correlations of the true population as closely as possible. The second condition implies that when we aggregate downscaled samples, the results need to be consistent with the original data. A more rigorous analysis is provided in the later section.

Synthetic data generation is often seen as a privacy-preserving way to combine multiple data sources in cases where direct collection is difficult or when common keys among multiple sources are missing. In applications where large-scale data collection involves manual surveys (e.g., demographics), or when the collected data is highly sensitive and cannot be fully released to the public (e.g., financial or health data), synthetically generated datasets become an ideal substitute. For example, due to privacy laws such as the General Data Protection Regulation \cite{eu-269-2014}, organizations across the world are forbidden to release personally identifiable data. As a result, such datasets are often anonymized and aggregated (such as geographical aggregation, where variables are summed or averaged across a certain region). Being able to join the lower-resolution sources, therefore, is a key step to reconstruct the full information from partial sources.

Common techniques for synthetic data generation are synthetic reconstruction (SR) \cite{beckman1996creating} and combinatorial optimization (CO) \cite{huang2001comparison,voas2000evaluation}. Existing approaches have specific data requirements and limitations which usually cannot be easily resolved. 

To address these challenges, we propose a new framework called \method (\textbf{Syn}thetic Data Generation via Gaussian \textbf{C}opula) to simulate microdata by sampling features in batches. The concept is motivated by \cite{jeong2016copula} and \cite{kao2012dependence}, which are purely based on copula and distribution fitting. The rationale behind our framework is that features can be segmented into distinct batches based on their correlations, which reduces the high dimensional problem into several sub-problems in lower dimensions. Feature dependency in high dimensions is hard to evaluate via common methods due to its complexity and computation requirements, and as such, Gaussian copula, a family of multivariate distributions that is capable of capturing dependencies among random variables, becomes an ideal candidate for the application. 

In this study, we make the following contributions:

\begin{enumerate}
    \item We propose a novel combination framework which, to the best of our knowledge, is the first published effort to combine state-of-the-art machine learning and statistical instruments (e.g., outlier detection, Gaussian copula, and predictive models) to synthesize multi source data.
    \item We perform simulation studies to varify \method's performance as a privacy-preserving algorithm and its ability to reproduce original datasets. 
    \item We demonstrate \method as a feature engineering tool, as well as an alternative to data collection in situations where gathering is difficult through a real-world datasets in the automotive the industry.
    \item We ensure the methodology is scalable at the production level and can easily incorporate new data sources without the need to retrain the entire model.
    \item To foster reproducibility and transparency, all code, figures and results are openly shared\footnote{See supplementary material available at \url{https://github.com/winstonll/SynC}}. The implementation is readily accessible to be adapted for similar use cases.

\end{enumerate}

\section{Related Works}
\subsection{Synthetic Reconstruction}
Synthetic reconstruction (SR) is the most commonly used technique to generate synthetic data. This approach reconstructs the desired distribution from survey data while constrained by the marginal distributions. Simulated individuals are sampled from a joint distribution which is estimated by an iterative process to form a synthetic population. Typical iterative procedures used to estimate the joint distribution are iterative proportional fitting (IPF) and matrix ranking. The IPF algorithm fits a n-dimensional contingency table base on sampled data and fixed marginal distributions. The inner cells are then scaled to match the given marginal distribution. The process is repeated until the entries converge. 
    
IPF has many advantages like maximizing entropy, minimizing discrimination information \cite{ireland1968contingency} and resulting in maximum likelihood estimator of the true contingency table \cite{little1991models}. However, IPF is only applicable to categorical variables. The \method framework incorporates predictive models to approximate each feature, which can be used to produce real-valued outputs as well and probability distribution that can be sampled from to produce discrete features. 
    
\subsection{Combinatorial Optimization}
Given a subset of individual data with features of interest, the motivation behind combinatorial optimization (CO) is to find the best combination of individuals that satisfy the marginal distributions while optimizing a fitness function \cite{barthelemy2013synthetic}. CO is typically initialized with a random subset of individuals, and individuals are swapped with a pool of candidates iteratively to increase the fitness of the group. Compared to SR based approaches, CO can reach more accurate approximations, but often at the expense of exponential growth of computational power \cite{wong2013optimizing}.

\subsection{Copula-Based Population Generation}
Copula is a statistical model used to understand the dependency structures among different distributions (details are discussed in Section \textit{Proposed Framework}), and has been widely used in synthetic data generation tasks \cite{kao2012dependence}. However, downscaling is not possible, and the sampled data stay at the same level of granularity as the input. Jeong et al. discuss an enhanced version of IPF where the fitting process is done via copula functions \cite{jeong2016copula}. Similar to IPF, this algorithm relies on the integrity of the input data, which, as discussed before, can be problematic in real-world settings. Our method, \method, is less restrictive on the initial condition of the input dataset as it only requires aggregated level data. Therefore \method is more accessible compared to previous approaches. 

\section{Proposed Framework} \label{proposed_method}

\subsection{Problem Description}
Throughout the rest of this paper, we assume that the input data comes from many different sources, $X_1,...,X_T$. Each source, called a batch, takes the form $X_t = [X_1, ..., X_M]$, where $X_m = [X_m^1, ..., X_m^{D_t}]$ is a $D_t$ dimensional vector representing input features of batch $t$. Together, there are $D$ features across the entire $B$ batches. Each $m \in M$ represents an aggregation unit containing $n_m$ individuals, and every individual belongs to exactly one $m$. All batches contain the same aggregation units, however, each has their own set of features. When referring to specific individuals within an aggregation unit, we use $x_{m,k}^d$ to denote the $d^{th}$ feature of the $k^{th}$ individual who belongs to the aggregation unit $m$. For every feature $d \in D$ and aggregation unit $m$, we only observe $X_m^d = \sum_{k=1}^{n_m} x_{m,k}^d/n_m$ at an aggregated level. We use the term \textit{coarse data} to refer to this type of low-resolution observations. In practice, aggregation units can be geopolitical regions, business units or other types of segmentation that make sense for specific industries. 

The goal of \method is to combine all batches in order to reconstruct the unobserved $\{x_{m,1}^d,\cdots, x_{m,n_m}^d\}$ for each feature $d$ in every batch and aggregation unit $m$. We assume that $M$ is sufficiently large, especially relative to the size of each $D_t$, so that fitting a reasonably sophisticated statistical or machine learning model is permissible, and we also assume that the aggregation units, $n_m$, are modest in size so that not too much information is lost from aggregation and reconstruction is still possible. Thus for a $M \times D$ dimensional coarse data $X$, the reconstruction should have dimensions $N \times D$, where $N = \sum_{m=1}^{M} n_m$ is the total number of individuals across all aggregation units. This finer-level reconstruction is referred to as the \textit{individual data}.

\method is designed to ensure the reconstruction satisfies the following three criteria \cite{munnich2003simulation}: 
\begin{enumerate}
    \item For each feature, the marginal distribution of the generated individual data should agree with the intuitive distribution one expects to observe in reality.
    \item Correlations among features of the individual data need to be logical and directionally consistent with the correlations of the coarse data.
    \item Aggregating generated individual data must agree with the original coarse data for each $m \in M$.
\end{enumerate}

\begin{figure*}[ht]
\centering
    \includegraphics[scale=0.4]{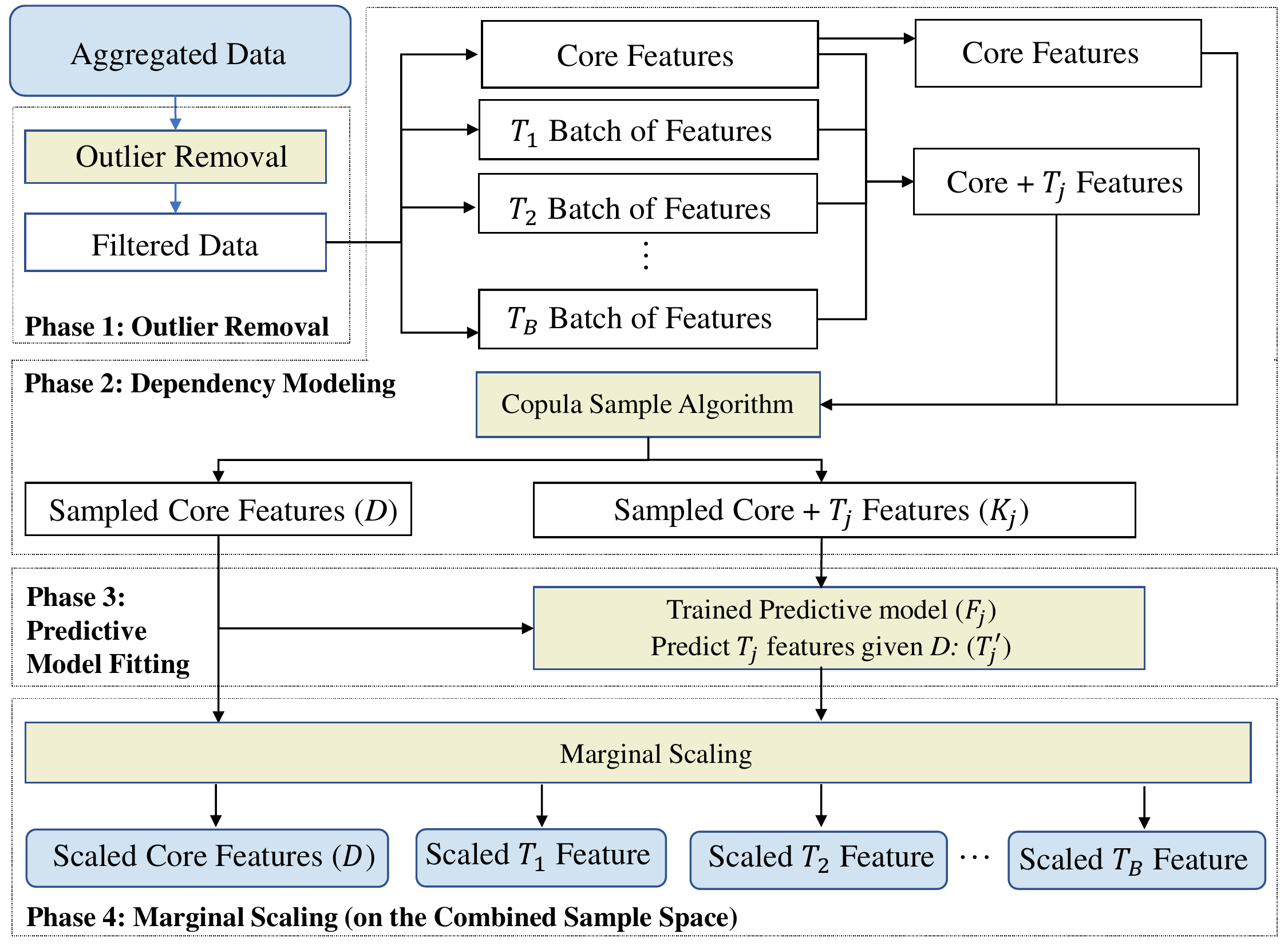}
\caption{\label{fig:flowchart}Flowchart of the \method framework} 
\end{figure*}

The main idea of \method is to simulate individuals by generating features in batches based on core characteristics and scaling the result to maintain hidden dependency and marginal constraints. As illustrated in Fig. \ref{fig:flowchart}, \method contains one preliminary phase and four key phases. Their formal descriptions are provided below.

\subsection{Phase 1: Outlier Removal} 
Outliers are the deviant samples from the general data distributions that may be a result of recording mistakes, incorrect responses (intentionally or unintentionally) or tabulation errors\cite{zhao2019lscp,li2020copod}. The presence of outliers can lead to unpredictable results \cite{zhao2019pyod,zhao2020suod}. Microsimulation tasks are sensitive to outliers \cite{passow2013adapting}, and conducting outlier detection before any analysis is therefore important. Outlier removal methods are often selected by the underlying assumptions and the actual condition of the data. Notably, the necessity of this process depends on the practitioners' definition of abnormality, although it is recommended in most cases. In Section \ref{sec:results_applications}, we study \method's performance with and without the Outlier Removal Step.
    
\subsection{Phase 2: Dependency Modeling} \label{Dependency Modeling}
We propose to use the copula model to address criteria i) and ii) since copula models, with their wide industrial applications, have been a popular choice for multivariate modeling especially when the underlying dependency structure is essential. First introduced by Sklar \cite{sklar1959fonctions}, a copula is a multivariate probability distribution where the marginal probability distribution of each variable is uniform. Let $X=(x_1, x_2...x_D)$ be a random vector in $\mathbb{R}^D$, and the marginal cumulative distribution function be \(P_i(x) = Pr[x_i<x]\), define \(U_i\) as 
\small
\begin{equation}
    U = (u_1, u_2, ..., u_D) = (P_1(x_1), P_2(x_2), ..., P_D(x_D))
\end{equation}
\normalsize

A copula of the random vector \(X\) is defined as the joint CDF of a random uniform vector U:
\small
\begin{equation}
    C(u'_1, u'_2, ... u'_D) = Pr(u_1<u'_1, u_2<u'_2, ..., u_D<u'_D)
\end{equation}
\normalsize

\RestyleAlgo{ruled}
\begin{algorithm}[!htp]
    \small
    \SetAlgoLined
    \KwData{Coarse Data}
    \KwResult{Simulated Individual Data}
    initialization\;
    X = input coarse data \\
    M = number of aggregated unit\\
    D = dimension of coarse data\\
    $\Sigma$ = $D \times D$ covariance matrix of $X$\\
    $\Phi$ = cumulative distribution function (CDF) of a standard normal distribution \\
    $F^{-1}_d$ =  inverse CDF of the marginal distribution of the $d^{th}$ component of $X$\\
    \For{$m$ in $1...M$}{
    Draw $Z_{m} = Z_m^1, \cdots, Z_m^D \sim N(0,\Sigma)$, where
    $N(\mu, \Sigma)$ denotes a $d$-dimensional 
    Normal distribution with
    mean $\mu$ and covariance matrix $\Sigma$ \\
        \For{$d$ in $1...D$}{
            $u_m^d = \Phi(Z_m^d)$\\
            $y_m^{'d} = F^{-1}_d(u_m^d)$ \\
            (This implies that $Y_m^{'d}$ follows the desired distribution) \\
            }
        \textbf{Return} $Y'_m = \{Y_m^{'i}\}_{i=1}^d$ 
        }
        \textbf{Return} $Y' = \{Y'_j\}_{j=1}^M$
    \caption{Gaussian copula sampling}
    \label{cop sample}
\end{algorithm}

In other words, we can describe the joint distribution of a random vector $X$ using its marginal distributions and some copula functions. Additionally, Sklar's Theorem states that for any set of random variables with continuous CDFs, there exists a unique copula as described above. It allows us to isolate the modeling of marginal distributions from their underlying dependencies. Sampling from copulae is widely used by empiricists to produced deserved multivariate samples based on a given correlation matrix and the desired marginal distributions of each of the components. Nelsen \cite{nelsen2007introduction} outlines a simpler version of Algorithm \ref{cop sample} for bivariate sampling, which can be generalized to multivariate cases.

In order to properly specify $F^{-1}_d$, we make a reasonable assumption that the size of each aggregation unit is significant and diverse enough such that $var(X_m^d)$ is approximately constant $\forall m$. This assumption implies that $var(X_m^d)$ can be estimated by $(\sum_{k=1}^{m} (X_m^d)^2 - \bar{X}_.^d)/M-1$, the unbiased sample variance estimator. Thus, given $\mu_m^d$ is observed from aggregation unit averages, $F^{-1}_d$ can be uniquely specified so long as we are willing to make an assumption on the distribution. For example, if the desired output is a positive continuous variable (such as income), we assume $F_{Y_m^d}(y)$ follows a lognormal distribution
with mean $\mu_m^d$ and standard deviation $\sigma_m^d$.
In the case of categorical variables, we assume $F_{Y_m^d}(y)$ follows a beta distribution with parameters $\alpha$ and $\beta$ such that $\frac{\alpha}{\alpha+\beta} = \mu_m^d$ and $\frac{\alpha\beta}{(\alpha+\beta)^2(\alpha+\beta+1)} = (\sigma_m^d)^2$. Our assumptions imply that $\mu_m^d$ and $\sigma_m^d$ can be derived from coarse data mean and standard deviation for feature $d$.

\small
\begin{equation}
    \mu_m^d = \textit{mean of feature $d$ in aggregation unit $m$}
\end{equation}
\begin{equation}
    \sigma_m^d = \sigma^d * \sqrt{M} * \sqrt{n_m}
\end{equation}
\normalsize
    
Algorithm 1 incorporates the above assumptions with Gaussian copula to satisfy criteria i) and ii).
    
\subsubsection{Gaussian Copula} \hfill\\
Being one of the most popular and simple form of copula models, the Gaussian Copula is easy to interpret, implement and sample (and will be the copula used in Section \ref{sec:results_applications}). It is constructed from multivariate normal distribution through probability integral transformation. For a given covariance matrix, $\Sigma$, Gaussian Copula requires the joint distribution of given variables can be expressed as:
\small
\begin{equation}
    C_{\Sigma}^{\text { Gauss }}(u)=\Phi_{\Sigma}\left(\Phi^{-1}\left(u_{1}\right), \dots, \Phi^{-1}\left(u_{d}\right)\right)
\end{equation}
\normalsize
where $\Phi$ is the CDF of normal distribution. 
\small
\begin{equation}
    V = \left({\Phi^{-1}\left(u_{1}\right)}, \ldots, {\Phi^{-1}\left(u_{d}\right)}\right)
\end{equation}
\normalsize
\small
\begin{equation}
     c_{\Sigma}^{\mathrm{Gauss}}(u)=\frac{1}{\sqrt{\operatorname{det} \Sigma}} \exp (-\frac{1}{2}\left(V \cdot\left(\Sigma^{-1}-I\right)\cdot V^{t} \right)\\
\end{equation} 
\normalsize
    
Because of the popularity of the normal distribution and the other advantages mentioned above, Gaussian copula has been one of the most widely adopted dependence models. The earliest application of Gaussian Copula was in the finance industry: Frees and Valdez applied the technique to insurance pricing \cite{frees1998understanding}, while Hull and White used it in credit derivative pricing \cite{hull2006valuing}. More recently, the application of Gaussian copula models have been found in many fields such as linkage analysis \cite{li2006quantitative}, functional disability data modeling\cite{dobra2011copula} and power demand stochastic modeling \cite{lojowska2012stochastic}. 

\subsubsection{Archimedean Copula}\hfill\\
Like the Guassian copula models, Archimedean copula modles are defined more generally as: 
\small
\begin{equation}
    C\left(u_{1}, \ldots, u_{p}\right)=\psi\left(\sum_{i=1}^{p} \psi^{-1}\left(u_{i}\right)\right)
\end{equation}
\normalsize
where $\psi :[0,1] \rightarrow[0, \infty)$ is a continuous, strictly decreasing and convex function \footnote{$(-1)^{k} d^{k} \psi(x) / d^{k} x \geq 0$ for all x $\geq$ 0 and $k=1$, \ldots, $p-2$,  and $(-1)^{p-2} \psi^{p-2} (x)$ \text { is non-increasing and convex.}}. 

With different $\psi$ functions, Archimedean Copula have many extensions. The popular cases are Clayton's Coupla which as been used in bivariate truncated data modelling \cite{wang2007analysis} and hazard scenarios and failure
probabilities modelling \cite{salvadori2016multivariate}, and Frank's Copula, which has been used in storm volume statistics analysis \cite{salvadori2004analytical} and drought frequency analysis \cite{wong2013comparison}.

Archimedean copulae have gained some attention in recent years and their unique generator functions have outstanding performances in particular situations. Yet the fitting process for the generator function can be computational expensive. As \method preforms data generation in batches and wishes to be computationally efficient, we will use Gaussian copula for modeling the dependency structure of features. 
    
\subsection{Phase 3: Predictive Model Fitting}
In theory, we could construct a complex enough covariance matrix with all variables, and then fit a giant copula model using Algorithm 1. However, two reasons make this approach nearly impossible: 1) the curse of dimensionality makes it difficult when $D$ is sufficiently large, and 2) certain columns of $X$ may update more frequently, and it is inefficient to train new models each time. In this section, we introduce an alternative method, Algorithm \ref{cop sample}, to resolve this issue with minimal computation requirements.

Starting with the core batch $T_0$, a subset of $X$ containing the utmost important features selected from \textbf{Phase 0}, we apply Algorithm \ref{cop sample} to $T_0$ and the resulting data set, $Y'$, should satisfy the first two criteria mentioned in \textbf{Phase 2}. 

Secondly, together with $T_0$, all $T_j$ are inputted into Algorithm \ref{batch sample} sequentially. The resulting sampled data set, $K_j$, contains features from $T_0$ and $T_j$. 
    
However, a problem immediately presents: the sampled individuals in $K_j$ do not match those from $Y'$. Furthermore, individuals sampled in each $K_j$ does not necessarily match the sampled individuals in $K_l$ (for $l \neq j$). To fix it, we can train a predictive model on $K_j$ to approximate the distribution of $P(T_j \mid S = s)$, and use the model to predict the values of features $K_j$ given their values of the core features from $X'$. The choice of the predictive model depends on the complexity and nature of the data on a case by case basis.
    
Finally, we merge the predicted values with the original data set $Y$ and iterate until all features are processed. This step is summarized in Algorithm \ref{batch sample}.

\RestyleAlgo{ruled}
\begin{algorithm}[!tp]
\small
\SetAlgoLined
\KwData{Coarse Data}
\KwResult{Simulated Individual Data}
    initialization\;
    $X$ = Input coarse data\\
    $B$ = total number of batches of non-core features\\
    $S$ = predefined set of core features\\
    $T_j$ = $j^{th}$ batch of non-core features\\
    $Y'$ = Initial sampled individual data with only core features using Algorithm \ref{cop sample}\\
    \For{$j$ in $1...B$}{
        $X_j' = X[S\cup T_j]$\\
        $K_j$ = Sampled data by applying Algorithm \ref{cop sample} to $X_j'$\\
        $F(T_j \mid S = s)$ = an approximated distribution function trained on $K_i$\\
        $Q_j$ = $\argmax_{Q}$ $F(Q\mid S = Y)$\\
        $Y'$ = $Y'\Join Q_j$ \text{where $\Join$ is the natural join operator}
        }
    \textbf{Return} $Y'$
    \caption{Batch sampling via Gaussian copula}
    \label{batch sample}
\end{algorithm}
    
\subsection{Phase 4: Marginal Scaling}
The final step is to address criterion iii), which is to ensure sampled individual data agree with the input coarse data.

If $Y^d$ is categorical with $c^d$ classes, we constrain the output, $Y'^d$ to the mean vector of $m$, $\mu_m^d = n_m \times X_m^d$. As $n_m$ is the population count of aggregation unit $m$, and $X_m^d$ is the observed proportion vector for feature $d$, $\mu$ is the count of each classes to be assigned to individuals for $m$. One thing to note that the predicted values from \textbf{Phase 3} for individual $k$ in aggregated unit $m$, represented by $p_{m,k}^d=\{p_{m, k, i}^d\}_{i=0}^c$, is a probability distribution. Hence it is natural to assume $Y^d \sim Multi(1, c^d, p_{m,k}^d)$, where $Multi(n, c^d, p^d)$ denotes a multinomial distribution with $n$ samples, each taking a category between $c^1, ..., c^d$, with probabilities $p^1, ..., p^d$. To determine the exact class of individual $k$, we generate a random sample from the distribution. After initial sampling, the percentage of each category may not match the marginal constraint. To resolve this, whenever a sample is produced, we subtract 1 from the corresponding dimension of the marginal distribution and resample if the corresponding dimension has already reached 0. The is summarized in Algorithm \ref{scaling}.

If $Y^d$ is continuous, the sampled mean and variance should be in proximity with the original coarse data given the way $F^{-1}_d$ is constructed in \textbf{Phase 2}. In case of small discrepancy, we apply the standard scaling of $Y' - (\mu_{sample} - \mu_{core})$ to horizontally shift each data point by the difference between sample mean and the coarse data mean. 

\RestyleAlgo{ruled}
\begin{algorithm}[!t]
\small
\SetAlgoLined
\KwData{Simulated Individual Data}
\KwResult{Individual Data with Categories Assigned}
    initialization\;
    $Y'$ = Initial sampled individual data outputted by Algorithm \ref{batch sample}. There should be $n_m$ individuals for each aggregation unit.\\
    $c^d = [c_1^d,...,c_k^d]$ categories for dimension $d$. \\
    $\mu_{m}$ = The marginal vector for aggregation unit $m$. \\
    \For{$d$ in $1...D$}{
        \For{$i$ in $1...m$}{
            \For{$j$ in $1...n_i$}{
                $p_i = Y'[j, :]$\\
                Draw class $\Tilde{c^d}$ from $Multi(1, c^d, p_j)$\\
                \textbf{If} $\mu_{i}[\Tilde{c^d}] > 0$ \\
                \textbf{Then} $Y_j = \Tilde{c^d}$ and $\mu_{i}[\Tilde{c^d}] = \mu_{i}[\Tilde{c^d}] - 1$ \\
                \textbf{Else} Repeat lines 9-11.\\
                }
            \textbf{Return} $Y_j = [Y_1,...,Y_{n_i}]$
            }
        \textbf{Return} $Y = [Y_1,...,Y_m]$
        }
    \caption{Marginal Scaling of Output Data}
    \label{scaling}
\end{algorithm}

\section{Results and Applications}
\label{sec:results_applications}
In this section, we demonstrate the validity of \method by a number of simulation studies, as well as show how \method can be used in real world applications. 

Through two demonstrations, we show:
\begin{enumerate}
    \item \method's reconstruction ability by measuring how close the generated dataset is to its original unaggregated version,
    \item the improvements on model accuracy when \method is used as a feature engineering tool when training data has limited number of features.
\end{enumerate}

For the first experiment, we use a dataset from a Canadian market research company with 65,000 respondents evenly selected across Canada. For the next two experiments, our data comes from the 2016 Canadian National Census, which is collected by Statistics Canada and compiled once every five years. The census is aggregated at the postal code level and made available to the general public. There are 793,815 residential postal codes in Canada (in the format L\#L\#L\#, where L is a letter and \# is a digit), with an average of 47 residents per postal code. The dataset contains more than 4,000 variables ranging from demographics, spending habits, financial assets, and social values. Table \ref{census data} illustrates a subset of this dataset with 3 postal codes and 4 variables. All datasets are made available in our GitHub repository.

\begin{table}[!tp]
    \resizebox{\columnwidth}{!}{  
        \begin{tabular}{c|c|c|c|c}
            \hline
            \textbf{Postal} & \# \textbf{Population} & \textbf{Avg Age} & \makecell{\textbf{\% with} \\ \textbf{Mortgage}} & 
            \makecell{\textbf{\% Speaks} \\ \textbf{two languages}} \\ 
            \hline
            \hline
            M5S3G2      & 467              & 35.1        & 0.32             & 0.69                       \\ \hline
            V3N1P5      & 269              & 37.2        & 0.35             & 0.67                       \\ \hline
            L5M6V9      & 41               & 49.1        & 0.67             & 0.43                       \\ \hline
        \end{tabular}
    }
    \caption{An excerpt of three variables from the census data}\label{census data}
\end{table}

\subsection{Reconstruction Accuracy Assessment}
To assess \method's performance as a privacy preserving algorithm, we run experiments by taking a dataset, identify an appropriate aggregation unit and group individual records to form proportions and/or averages. Then we try to reconstruct the original dataset by applying \method to the aggregated version. As the first of its kind, we found very little competing implementations of similar algorithms, and as such, we compare \method's performance against itself in different cases. Specifically, we study \method's performance with different aggregation unit sizes, as well as whether outlier removal (OR) was included.

\subsubsection{Data Description} \hfill\\
We use a dataset from a Canadian market research company with 65,000 respondents evenly selected across Canada. To keep the computations simple, we use 14 variables from this dataset, which are \textit{Age}, \textit{Gender}, \textit{Ethnicity}, \textit{Income}, \textbf{Education} and whether the respondent uses the internet to \textit{Read News}, \textit{Listen to Podcasts}, \textit{Sports}, \textit{Fashion}, \textit{Food}, \textit{Health}, \textit{Travel} and \textit{Social Media}. Forward Sortation Areas (FSA), which are geographical region in which all postal codes start with the same three characters, are used as aggregation units. 

\textit{Age} is a categorical variable with 7 classes, \textit{18 or under}, \textit{18-25}, \textit{26-34}, \textit{35-44}, \textit{45-54}, \textit{55-64} and \textit{65+}. \textit{Gender} is a binary variable, taking values of \textit{Male} and \textit{Female}. \textit{Ethnicity} assumes 5 classes (\textit{White}, \textit{Asian}, \textit{Middle Eastern}, \textit{African Canadian} and \textit{Others}. \textit{Income} has 13 classes (intervals of \$10,000 from \$0 up to \$100,000, \$100k-\$125k, \$125k-\$150k and \$150k+). Finally, \textit{Education} has 3 classes, which are \textit{High School or Below}, \textit{Post-secondary} and \textit{Postgraduate}.

All internet related variables are binary valued, with \textit{Yes} and \textit{No} as the only two possible values.

\subsubsection{Experiment Setup} \hfill\\
We first aggregate the data by FSA, and then reconstruct the original data using \method. After the generation, we rank all individuals by age and then gender, and accuracy is measured by comparing the number of matching cells from the raw data and the generated data with the same row number. As an example, the first individual in the raw data is a 20-24 year old female with \$40,000 - \$50,000 income, and uses the internet to read sports, listen to podcast and use social media. In the reconstructed data, the first individual is a 20-24 year old female but with \$50,000 - \$60,000 income and uses the internet to read sports, fashion and travel. In this case, \method's accuracy would be $6/12 = 50\%$, as we have correctly predicted \textit{age}, \textit{gender}, the consumption of \textit{sports} related content online, and have also correctly that she does not use internet for \textit{news}, \textit{food} and \textit{health} related content.

\subsubsection{Comparison by Aggregation Unit Size} \hfill\\
We run this analysis for all 65,000 individuals, and report our findings in Table \ref{accuracy_n}. In order to properly evaluate \method's performance, we give a breakdown of accuracy by the size of the aggregation unit, as well as whether or not the \textit{Outlier Removal Phase} was performed.

\begin{table}[!t]
    \resizebox{\columnwidth}{!}{ 
        \begin{tabular}{c|c|c|c|c|c}
        \hline
        \textbf{Agg. Unit Size} & 1-10 (n=45) & 10-25 (n=264) & 25-50 (n=509) & 50-100 (n=238) & 100+ (n=87) \\ \hline \hline
        \textbf{\% with OR} & 0.414 & 0.339 & \textbf{0.298} & \textbf{0.272} & 0.226 \\ \hline
        \textbf{\% wihtout OR} & \textbf{0.476} & \textbf{0.346} & 0.283 & 0.256 & \textbf{0.245} \\ \hline
        \end{tabular}
    }
    \caption{Reconstruction accuracy by size of aggregation unit}\label{accuracy_n}
\end{table}

We can see that reconstruction accuracy varies greatly. For smaller aggregation units, we can, on average expect close to 50\% accuracy in reconstruction, whereas for larger aggregation units, reconstruction accuracy is just over 20\%. This is expected because as the size of the aggregation unit grows, less information gets preserved from only observing averages or proportions of the categories. 

\subsubsection{Comparison by Variable Categories} \hfill\\
Another way to assess \method's performance is to look at accuracy across variables with different numbers of classes. In this particular application, we have 10 variables with 2 classes (\textit{Gender} and all 9 Internet related variables), 1 variable with 3 classes (\textit{Education}), 1 variable with 5 classes (\textit{Ethnicity}), 1 variable with 7 classes (\textit{Age}) and 1 variable with 13 classes (\textit{Income}). Reconstruction accuracy, grouped by variable category sizes, are summarized below, and benchmarked against the baseline measure of assigning by complete random.

\begin{table}[!tp]
    \resizebox{\columnwidth}{!}{ 
        \begin{tabular}{c|c|c|c|c|c}
        \hline
        \textbf{Number of Classes} & c=2 & c=3 & c=5 & c=7 & c=13 \\ \hline \hline
        \textbf{\% with OR} & \textbf{0.800} & \textbf{0.727} & 0.494 & 0.396 & 0.240 \\ \hline
        \textbf{\% wihtout OR} & 0.783 & 0.705 & \textbf{0.501} & \textbf{0.411} & \textbf{0.268} \\ \hline
        \textbf{Baseline} & 0.500 & 0.333 & 0.200 & 0.143 & 0.077 \\ \hline
        \end{tabular}
    }
    \caption{Reconstruction accuracy by number of categories}\label{accuracy_c}
\end{table}

One point worth mentioning is that the effect of including the outlier detection step varies. When fewer numbers of classes are present, the result suggests that outlier is an useful step to include, but when the number of classes increases, the effect of OR seems to diminish. 

\subsection{\method as a Feature Engineering Tool}
To assess the performance of \method as a feature engineering tool, we collaborate with a global automotive company (hereafter referred to as the "Client") that specializes in producing high-end cars to build a predictive model to better assist their sales team in identifying which of their current customers who have a leased vehicle are interested in buying the car in the next 6 months. This type of analysis is extremely important in marketing, as contacting customers can be both expensive (e.g., hiring sales agents to make calls) and dangerous (e.g., potentially leading to unhappy customers and therefore unsubscribing emails or services). Therefore building accurate propensity models can be extremely valuable.

Our experiment contains three steps. 1) We work with the Client to identify internal sales data and relevant customer profiles such as residential postal code, age, gender, 2) we apply \textit{probabilistic matching} to join sampled data together with the Client's internal data to produce an augmented version of the training data, and 3) we run five machine learning models on both the original and the augmented data, and evaluate the effectiveness of \method.

\subsubsection{Data Description} \hfill\\
There are 7,834 customers who currently lease a specific model of car from the Client in Canada. Our Client is interested in predicting who are more likely to make a purchase in the next 6 months. In Table \ref{auto leasing data} shown in the Appendix, we attach an excerpt of the Client's sale records. For security reasons, names and emails are removed. 

To augment this dataset, the Client selects 30 variables from the Census, which included information they do not know but could be useful features. The variables includes, demographics (personal and family), financial situations and how they commute to work. We include an excerpt of the sampled individual data obtained by apply \method to the Census in Table \ref{auto leasing data}; both datasets are available upon request.

\subsubsection{Probabilistic Matching} \hfill\\
A challenge of \method as a feature engineering tool is the fact that synthetic population is anonymous. In most applications, enterprise level data sources are individually identifiable through an identifier which may be unique to the company (e.g. customer ID for multiple products/divisions at one company) or multiple companies (e.g. cookie IDs which can be shared between multiple apps/websites). This makes merging different data sources very easy as the identifier can be used as the primary key. By construct, however, synthetic population individuals are model generated from anonymous data, and therefore cannot be joined by traditional means. Here we present an alternative method called \textit{Probabilistic Matching}.

Because \method produces anonymous information (i.e. data that cannot be attributed to a specific person, such as name, email address or ID), we use age, gender, ethnicity, and profession as good identifiers to the real population. In table 3 we show the first few customers supplied by our industry partner. We also provide the list of synthetically generated persons for postal code V3N1P5 in Appendix Table \ref{V3N1P5}, and use this as an example to demonstrate how probabilistic matching can be done on the first customer.

Client data show that a 53 year old male, who lives in the area of V3N1P5 made a lease purchase. In this case, we have three indicative measurement for this customer - this buyer is \textit{53 years old}, \textit{male} and lives in \textit{V3N1P5} (an area in Burnaby, British Columbia, Canada). In our synthetic data, the closet match would be the tenth person, as two of the three indicators (postal code and genders) match precisely, and the last indicators matches closely (age of 54 vs. 53, which is a difference of 1 year). We can conclude that this customer, who leased an SUV of model type 3, is likely to be ethnically Chinese, an immigrant with a bachelor's degree, an income of between \$90k to \$99k and speaks 2 languages.

\subsubsection{Method Evaluation} \hfill\\
We train 5 different classifiers on both the partner's data, as well as the augmented dataset to predict whether a customer buys the leased vehicle. We train Logistic Regression (LR), Decision Tree (DT) \cite{hu2019optimal}, Random Forest with 500 tress (RF) \cite{breiman2001random}, SVM with Radial Basis Kernel and 2-Layer Neural Network (NN). Standard grid search cross validation is used to ensure that the best hyperparameters are selected, and the models' performances are summarized in Table \ref{performance_classifier}. 

\begin{table}[!tp]
    \resizebox{\columnwidth}{!}{
    \begin{tabular}{c|c|c|c|c|c}
    \hline
     & \textbf{LR} & \textbf{DT} & \textbf{RF} & \textbf{SVM} & \textbf{NN} \\ 
     \hline \hline
    \textbf{Original Data} & 0.615 & 0.639 & 0.704 & 0.693 & 0.688 \\ 
    \hline
    \textbf{Augmented Data} & \textbf{0.662} & \textbf{0.711} & \textbf{0.730} & \textbf{0.806} & \textbf{0.739} \\
    \hline 
    \end{tabular}}
    \caption{Comparisons of accuracy measures of 5 different classifiers trained on original and synthetic population augmented data}
    \label{performance_classifier}
\end{table}

In all five cases, augmented data produces a higher classification accuracy than the raw data from our industry parnter. Accuracy increases range from slightly over 2.5\% (RF) to as much as 11\% (SVM), with an average increase of 6.2\%. This increase is both technically significant, as well as practically meaningful, as the Client would easily apply this model to their business and achieve grow their sales. 

This case study has shown that Synthetic Population is an effective way to engineer additional features to situations where the original training data is limited. As explained in early sections, \method takes coarse datasets and generate estimates of individuals that are likely to reside within the given postal code area. Although \method does not produce real people, the generated "synthetic" residents both closely resembles the behaviours of true population and is also consistent with the available sources. We demonstrate that it is a viable data augmentation technique.

\section{Conclusion and Future Directions}
In this work, we propose a novel framework, \method, for generating individual level data from aggregated data sources, using state-of-the-art machine learning and statistical methods. To show the proposed framework's effectiveness and boost reproducibility, we provide the code and data used in simulation studies described in Section \ref{sec:results_applications}. We also present a real-world business use case to demonstrate its data augmentation capabilities. 

As a first attempt to formalize the problem, we see multiple areas where future works can improve upon. First of all, our method relies on Gaussian copulae and this can be further extended by leveraging other families of copula models to better model the underlying dependency structures. Secondly, we use beta and log-normal distributions to approximate marginal distributions for categorical and continuous variables, respectively, and other families of distributions could be considered (e.g., the $\kappa$-generalized model \cite{clementi2016kappa} can be used for money related distributions). Lastly, a better similarity metric can be designed to assess generated data against its original input.

\bibliographystyle{./bibliography/IEEEtran}
\bibliography{./bibliography/sync}

\section*{APPENDIX}
\newpage

\begin{sidewaystable}
    \centering
        \begin{adjustbox}{max width=0.95\textheight}
        \begin{tabular}{c|c|c|c|c|c|c|c|c|c|c}
        \hline
        \textbf{Postal} &  
        \textbf{Sex} & 
        \textbf{Age} & 
        \textbf{Ethnicity} &
        \makecell{\textbf{Immigration} \\ \textbf{Status}}  &
        \textbf{Education} &\textbf{Profession} &
        \makecell{\textbf{Marital} \\ \textbf{Status}} &
        \textbf{Family Size}   &
        \textbf{Income}  &
        \makecell{\textbf{Languages} \\ \textbf{Spoken}}\\ 
        \hline
        \hline
        V3N1P5   & F      & 19   & Latin &Immigrants & No degree & Ed services &Married & 5+ & $<$\$10k & 1 \\\hline
        V3N1P5   & F      & 65+  & Chinese & Immigrants & No degree &Food services &Widowed & 3 & \$10k to \$19k & 2\\\hline
        V3N1P5   & M      & 51   & Korean & Immigrants & College &Waste management &Separated & 1 & $<$\$10k & 2   \\\hline
        V3N1P5   & M      & 60   & Korean & Immigrants & Master &Finance & Married & 2 & $<$\$10k & 2        \\\hline
        \hline
        \end{tabular}
        \end{adjustbox}
        \caption{An excerpt of simulated data for one postal region}\label{V3N1P5}
    
    \vspace{1\baselineskip}
    \centering
        \begin{adjustbox}{max width=0.95\textheight}
        \begin{tabular}{c|c|c|c|c|c|c|c|c|c|c|c|c|c|c}        
        \hline
        \makecell{\textbf{PostalCode}}& 
        \makecell{\textbf{Gender}}& 
        \makecell{\textbf{PersonAge}} & 
        \makecell{\textbf{old car VOI}} & 
        \makecell{\textbf{Dealer where } \\ \textbf{old car was} \\ \textbf{purchased}}& 
        \makecell{\textbf{date first} \\ \textbf{email sent}}& 
        \makecell{\textbf{date last} \\ \textbf{email sent}}& 
        \makecell{\textbf{unsubed while} \\ \textbf{in LYOL} \\ \textbf{ flag Y/N}}& 
        \makecell{\textbf{Finished Full}\\\textbf{Cadence}}& 
        \makecell{\textbf{Lease }\\ \textbf{terminated} \\\textbf{while in LYOL}}& 
        \makecell{\textbf{Sold while}\\\textbf{ in LYOL}}& 
        \makecell{\textbf{Lease renewed}\\\textbf{flag Y/N}} & 
        \makecell{\textbf{new Purchase}\\ \textbf{Date}}& 
        \makecell{\textbf{new car}\\ \textbf{signature}}& 
        \makecell{\textbf{dealer where} \\ \textbf{new car was}\textbf{purchased}}\\
        \hline \hline
        V3N1P5           & M      & 53        & XXX SUV 3   & Brian Jessel Dealership            & 4/9/2018              & 10/18/2018           & N                              & Y                     &                                &                    &                        &                   &                   &                                    \\
        H7T1T4           & M      & 68        & XXX SUV 4   & XXX Dealership in Laval            & 2/20/2018             & 6/20/2018            & N                              &                       & Y                              &                    & Y                      & 7/10/2018         & Lease             & XXX Laval                          \\

        L9X0S4           & M      & 45        & XXX Sedan 3 & Georgian XXX                       & 5/6/2019              & 7/8/2019             & N                              &                       &                                &                    &                        &                   &                   &                                    \\
        H9B1A5           & F      & 45        & XXX Sedan 3 & XXX Canbec                         & 1/17/2019             & 7/24/2019            & N                              & Y                     &                                &                    &                        &                   &                   &                                    \\
        L4S1W5           & M      & 52        & XXX SUV 3   & XXX Autohaus                       & 3/29/2018             & 10/11/2018           & N                              & Y                     &                                &                    & Y                      & 10/31/2018        & XXX SUV 3         & Lease                              \\
        H7P0B9           & F      & 43        & XXX Sedan 3 & XXX Laval                          & 4/1/2019              & 6/25/2019            & N                              &                       &                                &                    &                        &                   &                   &                                   \\
    \hline \hline
    \end{tabular}
    \end{adjustbox}
    \caption{An excerpt of auto leasing data}\label{auto leasing data}

\end{sidewaystable}
\end{document}